# PROTOTYPING AND PERFORMANCE ANALYSIS OF A QoS MAC LAYER FOR INDUSTRIAL WIRELESS NETWORK

**Adrien van den Bossche, Thierry Val, Eric Campo**

*University of Toulouse, LATTIS EA4155, SCSF group,*
*IUT de Blagnac-UTM BP60073, 1 pl. G. Brassens, 31703 Blagnac, France*
*{vandenbo, val, campo}@iut-blagnac.fr*

Abstract: Today's industrial sensor networks require strong reliability and guarantees on messages delivery. These needs are even more important in real time applications like control/command, such as robotic wireless communications where strong temporal constraints are critical. For these reasons, classical random-based Medium Access Control (MAC) protocols present a non-null frame collision probability. In this paper we present an original full deterministic MAC-layer for industrial wireless network and its performance evaluation thanks to the development of a material prototype.  Copyright © 2002 IFAC

## 1. INTRODUCTION

Today, a typical wireless ad-hoc network technology has to propose strong and reliable mechanisms for each level of the OSI model: Physical layer (PHY) must deal with low Bit Error Rate, Medium Access Control layer (MAC) must avoid collisions and solve hidden terminal, Network layer (NWK) must enable automatic routing and insure reliability for mobile nodes, and so on. For an industrial application, a higher reliability is required: communication technology must propose some guaranties depending on the application (temporal bounding on transmission latency and packet forwarding, minimal throughput, and maximal packet loses…). For all these reasons, adding Quality of Service (QoS) functionalities to the network technology is mandatory in real-time monitoring sensors network application.

There are numerous NWK-level QoS protocols for wire networks like *IntServ* (Wroclawski, 1997), *DiffServ* (Nichols *et al.* 1998a, b) or for wireless mobile ad-hoc network like QOLSR (Badis *et* Al Agha, 2004). These QoS techniques are generally based on a traffic admission control system: if the network capacity is lower than the requirements of the candidate traffic, network refuses to handle this traffic. Nevertheless, the traffic admission technique needs 1) a fine description of each traffic potentially handled by the network and 2) an exhaustive knowledge of the communication resources, which is not simple in the case of a wireless network where PHY and MAC behavior is difficult to predict. Ideally, on a wireless technology implementing QoS, the MAC-layer should be able to not only send/receive traffic with a high level of guarantee but also to return information on medium capacity in order to help the relevance of the traffic control system at NWK-level.

The aim of this paper is to present an original MAC-layer for industrial applications based on IEEE 802.15.4 LP-WPAN with QoS implementing a full deterministic medium access method. Thru the sections of this paper, we first present typical requirements for an industrial application of wireless sensor network. Then we present an overview of the IEEE 802.15.4 standard and discuss weaknesses we identified. We then propose a new, totally deterministic medium access method and its performance evaluation by a material prototyping.

## 2. INDUSTRIAL AND ROBOTIC APPLICATION REQUIREMENTS

Considering an industrial sensor network application, random network congestions are not acceptable: a typical sensor application in an industrial environment primarily needs a high level of reliability. Messages must be delivered in time, without error with a preliminary QoS negotiation to define minimal bandwidth, maximal latency on message delivery and maximal message losses.

To illustrate the typical communication needs and an example of network topology, we chose a mobile robotic application where robots can communicate together (*cooperating robotic application*). The wireless network allows just as well sensors / actuators communications (intra-robot communications) as cooperating messages exchanges (inter-robots communication). The network topology is illustrated on fig. 1. For this type of application, transmitted data impose hard temporal constraints: for example, sensors like ultrasound sonar (obstacle sensors) cannot accept variable transmission delay due to collisions and retransmissions. While any transmission technology using radiofrequency medium depends on an imperfect medium, the MAC-layer has to resolve medium access issues without introducing random parameters.

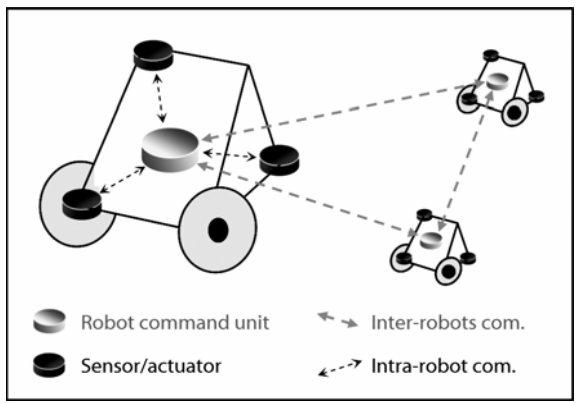

Fig. 1. Typical network topology for intra-robot and extra-robot communications cohabitation.

Finally, a typical industrial sensor application requires a reliable and energy-saver communication system. Average throughput is quite small – 100kbps at most – a higher reliability is preferable than a higher throughput. IEEE has recently introduced a LP-WPAN (Low-Power Wireless Personal Area Network) taking into account the constraints mentioned previously. This LP-WPAN standard is IEEE 802.15.4. In addition, the availability and low-cost of these devices are real advantages for the design of prototypes. The works of the IEEE 802.15.4 task group whose principal characteristics are detailed in the following section are the bases of our developments.

## 3. PRESENTATION OF IEEE 802.15.4 TECHNOLOGY

IEEE 802.15.4 standard (IEEE 2003) proposes an original two-layer protocol stack (physical-layer and data link-layer) for low power transceivers and low baud rate communications between embedded devices. Innovative concepts optimize energy saving. Moreover, IEEE 802.15.4 standard is promoted by the ZigBee Alliance (ZigBee Alliance, 2005) as the physical-layer and data-link-layer of the ZigBee Network specifications.

### 3.1 Overview

IEEE 802.15.4 proposes two PHY layers: PHY868/915 and PHY2450. The first one operates on both 868MHz and 915MHz radio bands. It proposes a very low data-rate (20kbps at 868MHz and 40kbps at 915MHz) with a simple BPSK modulation. The PHY2450 layer is more interesting: it allows a greater throughput (250kbps) thanks to an O-QPSK modulation. Moreover thanks to its Direct Sequence Spread Spectrum (DSSS) coding, PHY2450 has excellent noise immunity (IEEE, 2003). The two PHY layers were designed for maximum energy saving: protocols are optimized for short and periodical data transfers. Nodes mostly stay in a "sleeping" mode called doze mode. Radio modem allows ultra low power consumption (40μA) (Freescale Semiconductors, 2005) and nodes become operational in a very short time (330μs). In doze mode, all radio functionalities are switched off, removing the ability to receive messages. The waking time has to be set before going in doze mode (synchronous wake-up) but sleeping devices may also wake-up if a local event occurs (asynchronous wake-up): motion detection for example.

### 3.2 Medium Access Control (MAC) and topologies

The standard IEEE 802.15.4 proposes two data link-layer topologies: Peer-to-Peer and Star. Peer-to-peer topology makes possible direct data transfers between devices in radio range on the same radio channel. Access to the medium uses the CSMA/CA protocol without RTS/CTS mechanism. On the contrary, Star topology needs a star coordinator: all data transfers go through the coordinator and messages are buffered during the dozing period. This functionality is called indirect data transfer. Star topology allows high energy saving thanks to an optimal distribution of sleeping periods between embedded devices. For synchronization, the star coordinator sends beacon frames. Inter-beacon period is called superframe. During the superframe, the nodes sleep until the next beacon, wake up and receive the beacon, ask the star coordinator for pending data, transmit and receive and then go to doze mode again.

In addition to the classical CSMA/CA-based medium access method, IEEE 802.15.4 proposes a Contention Free method for the Star beaconed topology. Nodes can request for Guaranteed Time Slots (GTS) to the star coordinator. A GTS consists in one or several time slots dedicated to a particular node and cannot be used by other nodes. GTSs are announced by the beacon frame, a superframe contains up to seven GTS. The number of GTS reservations for a terminal node is directly linked to its communication bandwidth. This process of medium access reservation provides Quality of Service properties like bandwidth reservation or latency guarantees (Huang *et al.* 2006), like 802.11e HCF (IEEE, 2005).

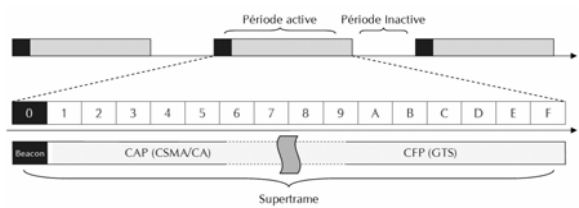

Fig. 2. IEEE 802.15.4 superframe structure.

The IEEE 802.15.4 superframe mixed structure (Fig. 2) combines both methods as follows: First, a star coordinator sends a beacon frame to indicate the network and coordinator addresses, the nodal data pending, the sizes of the Contention Access Period (CAP) and Contention Free Period (CFP). Then starts the CAP where the nodes and coordinator send/receive frames using CSMA/CA protocol. This time is also used for request from a node to obtain GTS in the next superframe. At the end of the CAP, the CFP starts as defined by the coordinator and broadcasted by the beacon. Medium access is possible only if the node has successfully obtained a GTS. At the end of the CFP, all nodes go to doze mode if not already and wait until the next beacon scheduled broadcast by using an internal wake up timer. This sleeping period is optional but greatly advised for energy saving.

$$BI = 15.36\,ms * 2^{BO} \text{ with } 0 \leq BO \leq 14 \quad (1)$$

$$SFAP = 15.36\,ms * 2^{SO} \text{ with } 0 \leq SO \leq BO \quad (2)$$

Therefore, the superframe is characterized by two temporal parameters Beacon Order (BO), Superframe Order (SO) announced in beacon frames: BO defines the time interval between two beacon messages. Beacon Interval (BI) is calculated as mentioned in (1). SO defines the SuperFrame "Active Portion" (SFAP = TCAP+TCFP) and is calculated as mentioned in (2). If BO and SO values are small, the network is more reactive (low latency) with lower energy saving. The greater is the difference between BO and SO, the more energy is saved. Thus, an appropriate value for these two parameters will be required considering the applications requirements.

## 4. IDENTIFIED PROBLEMS ON MAC LAYER

As shown in the above section, 802.15.4 adopts an interesting mechanism of medium access reservation (GTS) to make free some privileged nodes from the collision phenomenon. The medium reservation is conditioned by two factors: First, the network must be maintained within its capacity and avoid saturation (Misic *et al.* 2006). Unfortunately, the standard does not grant to a star coordinator the capability to permanently maintain some bandwidth for a particular node. The GTS reservation process works as "first come, first served" and is not an acceptable rule of distribution in terms of Quality of Service. Second, the primitive call "GTS.request" generates a message sent to the star coordinator during the CAP using the CSMA/CA protocol. As this protocol is Best-effort, it can not provide any temporal guarantee. By extension, the primitive call "association.request" message for joining a network is sent by using the same protocol and is therefore also not temporally guaranteed.

To achieve a communication network between sensors in an application with temporal constraints, it is essential to insure bandwidth and network latency for a number of known critical nodes. Moreover, sensors may have different communication requirements: strong sporadic flows, regular flows with time dependency, etc.

The standard IEEE 802.15.4 has other weaknesses:

- A connected node cannot preserve its GTS lease. To renew the GTS, a new request must be sent during the CAP. The possibility to request an extension of GTS allocation should be available during the allocated GTS.
- The GTS frequency is based upon the superframe frequency, the star coordinator BO and SO internal clock parameters. The nodes may only need to communicate from time to time and not on a regular basis. In other words, it is extremely difficult for sensors with different data communication needs to cohabit on the same star without loss of continuity and optimal GTS distribution.
- If several stars are in the same radio range and on the same channel, there is a high probability for collisions even during the CFP because the standard does not provide communication protocols between star coordinators.

According to all these observations, the mechanism of medium reservation could be greatly improved by:

- A fully deterministic access method to insure GTS for selected known nodes at each superframe,
- A more flexible GTS allocation to support various access frequencies and bandwidth,
- The introduction of a new protocol between star coordinators in order to avoid GTS collisions.

# 5. A NEW FULL DETERMINISTIC MAC LAYER

In the above section, we exposed some difficulties of the actual IEEE 802.15.4 standard. In this section, we present an original MAC layer implementing a fully deterministic medium access method. The goal is to reinforce the GTS mechanism and increase flexibility for the medium reservation and a communication set between stars.

## 5.1 New proposed functionalities

The proposed mechanisms are intended to achieve the following new functionalities:

- With the present standard, only nodes can request for a GTS. We propose to give a star coordinator the ability to allocate GTS at any time to any known node, in anticipation of a request. This functionality makes possible deterministic network associations for critical nodes. We call this ability PDS, for Previously Dedicated Slot.
- With the present standard, the GTSs were managed by the coordinator with internal messages within the star. We now propose a mechanism to extend these communications between coordinators to avoid "GTS collisions" (two coordinators give a same GTS for two nodes by two different stars in the same radio range).
- With the present standard, the GTSs were placed in the CFP, at the end of the superframe. We propose that the GTSs will be laid out anywhere in the superframe at the coordinator discretion. This functionality will enable us to optimize GTS distribution and with a possible extension to generate an optimized global superframe composed of several stars.
- With the present standard, a GTS appeared in every single superframe after allocation. We now suggest regulating this GTS inclusion in the superframe at a lower frequency to fit the node needs. Thus, a GTS can appear in one superframe out of two, one out of four, one out of eight, etc. We introduce the notion of reservation level $n$, an integer from 0 to $n_{MAX}$. The GTS of a node with a reservation level n will appear in every $p = 2^n$ superframes. It will allow different QoS traffics to cohabit within the same star without need for adjustments of BO and SO parameters.
- With the present standard, GTSs were allocated for a limited time and the lease could only be maintained via a repetition of renewed GTS request and therefore the continuity was jeopardized. We propose that GTSs will be allocated for an unlimited time, unless a release request by the node or a release notification by the star coordinator is issued (inactivity timeout, for example).

## 5.2 Stars cohabitation on a common radio range

If several coordinators cohabit in close proximity on the same radio channel, the GTS attribution must be decided in agreement with the other coordinators. In the specification, ZigBee describes a special node called PAN Coordinator. We propose to centralize all GTS requests messages to this node. Thus, the PAN coordinator can manage GTS repartition and ensures there is no GTS collision. Moreover, as GTS can appear in every $p = 2^n$ superframes, the decision of GTS attribution must be taken regarding not only the instantaneous network load, but also the future GTS. This decision is easier if only one device can take it, with an exhaustive vision of $2^{n_{MAX}}$ superframes. Typically, a GTS request message is transmitted by a node to its coordinator; the coordinator relays this request to the global coordinator and receives in return an authorization of allocate the GTS with a reservation level $n_{GTS}$.

## 5.3 Temporal organization of the beacons

Another problem in the actual 802.15.4 standard is the collision of beacons. IEEE 802.15.4b tasking group (IEEE, 2006) is working on this problem but 802.15.4b does not propose a deterministic way to avoid beacon collisions. Our solution proposes that PAN coordinator regulates the beacons like GTSs by distributing some specific timeslots dedicated to beacon frames: we call this beacon GTS Guaranteed Beacon Slot (GBS). GBS and GTS information are broadcasted in PAN coordinator beacons called superbeacons: this solution also solves the "hidden coordinator" problem. Like GTS, a GBS can appear every $p = 2^n$ superframes depending of the latency needed by the star. Star coordinators act as nodes to the PAN coordinator and have GBS with a reservation level $n_{GBS}$.

## 5.4 Conclusion on the MAC method proposed

The MAC method proposed permits to make full deterministic accesses to the medium. Nevertheless, contention access is still possible by using CAP timeslots (CSMA/CA). We can consider that determinist access GTS, GBS and PDS ensure minimal medium accesses and provide minimal bandwidth and maximal latency bounds. Of course, higher bandwidth and smaller latency can be attempted, but without guarantee.

# 6. VALIDATION BY PROTOTYPING

The new evoked functionalities need to be tested in order to validate the new medium access method. We proceed to several studies, first by simulation with the design and development of an original simulation tool and then by *Petri nets* (formal and mathematical validation). Another way to validate our proposition is prototyping, which is the main topic of this paper – simulation and *Petri nets* studies are presented in some other submitted papers (van den Bossche, A. *et al*, 2007). If simulation generally enables to get some good results on performances and scalability of the protocol, some functionalities such as signal propagation or antennas characterization are easier to test via prototyping. Moreover, final performances on throughput or delay may be different from simulation results because of the PHY layer which is real. For those reasons, performances characterization via prototyping seems to be fundamental.

*6.1 Presentation and characterization of the designed platform*

Our test bench is based on Freescale[TM] IEEE 802.15.4 / ZigBee solution. It consists in a dual chip module: standard 8 bit microcontroller MC9S08 and a specific 2.4GHz radio modem compatible with IEEE 802.15.4 physical layer specifications. Freescale[TM] IEEE 802.15.4 solution presents a real advantage: the MAC level is totally reprogrammable in C-language, which enable us to modify the standard medium access method to fit with our proposition and enabling real prototype performance measurement. Our prototype network is made of some Freescale[TM] 13192-SARD cards (top and right on fig. 3) and some cards developed in laboratory (left on fig. 3) based on Freescale[TM] ZRD-01 module. All measures are realized in an anechoic chamber at short distance (max 2 meters) in order to avoid performance degradation due to transmission errors on radio. Measures results are sent to a computer via RS232 serial port (DB9 connector).

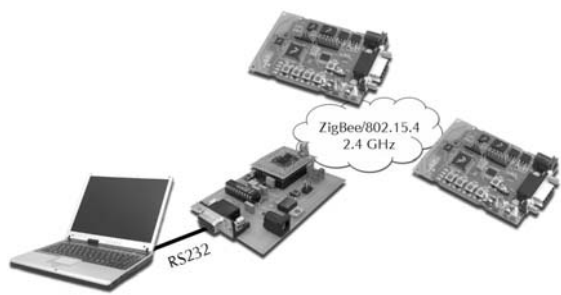

Fig. 3. IEEE 802.15.4 cards used for prototyping.

Before getting performance of the proposed MAC layer, we first characterized our platform in a raw context, without MAC implementation. The goal of this preliminary study is to obtain performance of the hardware platform and the IEEE 802.15.4 physical layer.

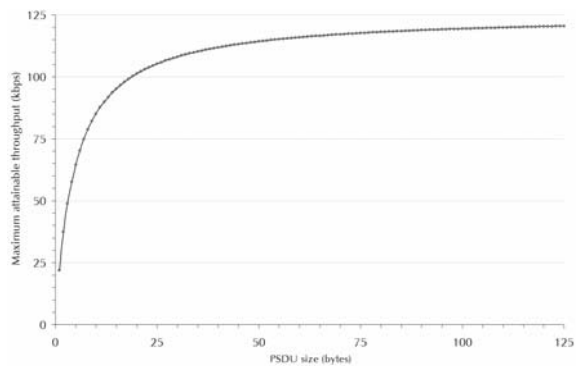

Fig. 4. Maximum attainable throughput according PSDU size without medium access method and acknowledges.

Fig. 4 represents maximum attainable throughput according PSDU (*PHY Service Data Unit*) size without medium access method. These results were obtained by using two 13192-SARD modules: the first sends frames with different packet size and the second one is blocked in receive mode. The data transmission is realized without acknowledgement while frames containing errors are simply ignored by the receiver. As shown on fig. 4, maximum throughput is logically obtained when PSDU size is set to the maximum (127 bytes) and is closed to 120kbps, which is very far from the 250kbps (theorical on *baseband*). Considering that there is no MAC layer and acknowledges, practical throughput may be close to baseband throughput. In fact, this poor practical throughput is due to an important inter-frame delay imposed by the *packet mode* protocol used on *SPI-bus* (*Serial Peripheral Interface*) between the microcontroller and the radio modem as illustrated in fig. 5 which represents total transmission delay including SPI *and* radio transmission.

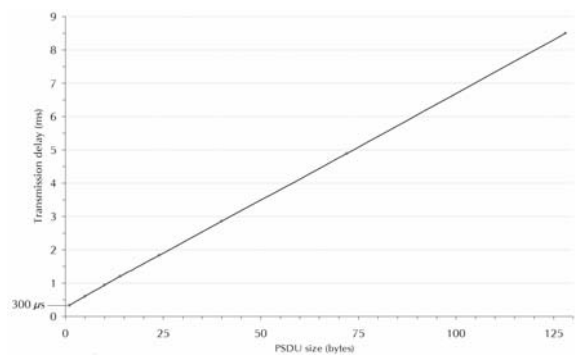

Fig. 5. Total transmission delay according PSDU size, including SPI-bus and radio delays.

In future, this point should be improved by using another protocol on SPI-bus called *stream mode* which enables direct sending of data packet in real-time. In fact, this optimization part is not the topic of this paper and will be mentioned as a potential perspective. Nevertheless we notice that all the performance evaluations presented in this paper are based on the *packet mode*.

*6.2 Determination of guaranteed throughput according to reservation level and BO*

The goal of this study is to determine the maximum attainable throughput of a communication using the proposed MAC protocol according to the temporal parameter BO (*Beacon Order*) and the GTS reservation level $n$.

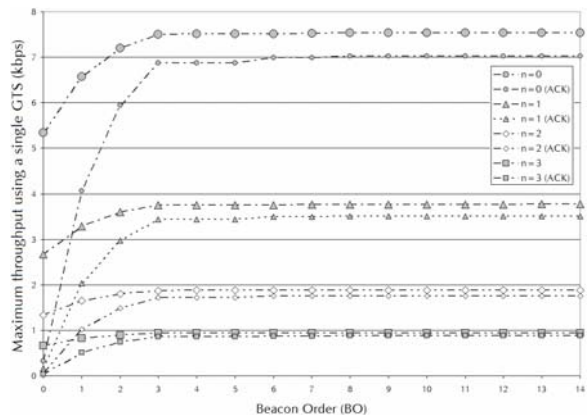

Fig. 6. Maximum throughput for a communication using a single GTS.

Fig. 6 represents the maximum attainable throughput of a unidirectional communication using one IEEE 802.15.4 slot (GTS) with / without acknowledgement frame, while medium access is done via the proposed deterministic medium access method. We can see on fig. 6 that the lowest *BO* value is unusable because in this case, timeslots are too short to allow transmitting DATA frames and an ACK frame. This study enables us to consider a minimum *BO* value of $BO = 1$ (*superframe duration SD* = 30 ms). For $BO = 0$, a time slot is not long enough to contain one DATA and one ACK frame. For $BO$ = 1 or 2, timeslots are too short to contain the maximum size of an 802.15.4 frame so throughput increases with *BO*. For *BO* values of 3 or more, throughput is constant. Nevertheless, while it is not the topic of this study, we note that high *BO* values increase delay of deterministic accesses.

*6.3 Maximal guaranteed throughput* (*deterministic aspect*)

This last performance study proves the deterministic aspect of the presented MAC method. In fact, while CSMA/CA based MAC have poor performance when the number of stations increases, the MAC method we propose guaranties the single GTS throughput even if the number of stations is important, as we can see on fig. 7. In fact, all stations have their own GTS so medium access is still guaranteed, even in a dense traffic context.

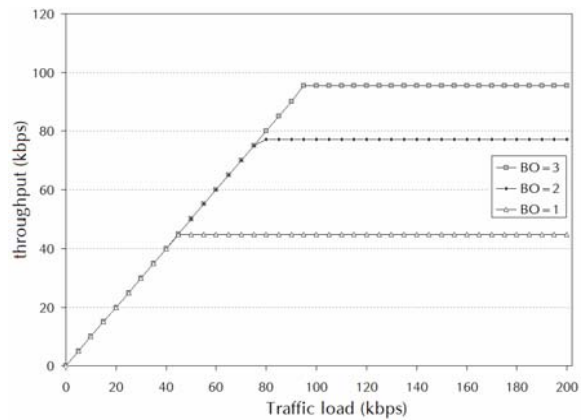

Fig. 7. Usual throughput in function of global traffic load on the entire network.

Results of fig. 7 can be compared with the classical bell curb of CSMA/CA (IEEE 802.15.4 CSMA/CA MAC-layer simulation in Gang, L. *et al*, 2004) where global network performances usually collapse if network solicitation is important. With the proposed MAC-layer, the MAC-level throughput is guaranteed whenever medium solicitation is important. We can conclude on the real deterministic aspect of the proposed MAC method.

## 7. CONCLUSION AND FUTURE WORKS

In this paper, we have presented the IEEE 802.15.4 technology and identified some gaps on MAC-layer concerning Guaranteed Time Slots. To solve theses problems, we have proposed an original MAC-layer for industrial and robotic wireless network based on a full deterministic medium access. This new MAC has been validated by complementary methods, while only the real prototyping is presented in this paper. Results are interesting: the proposed medium access method presents good performance considering throughput, especially if medium solicitation is important (high traffic load) whereas CSMA/CA protocol has poor performance on high traffic load. Moreover, the introduction of the *n* parameter (GTS reservation level) permits different physiognomy traffics to be carried over the network. Thanks to this new MAC-layer, industrial sensors applications like control/command can be considered with LP-WPANs.

Works in progress are numerous and are focussed on message transport latency: while we prove in this paper that a MAC-level throughput can be guaranteed, it could be interesting to also guarantee transmission delay (i.e. medium access period). We also note in this paper that internal *SPI-bus* protocol may be improved in order to get better temporal performances on data-packet processing. Another future perspective is the study of the energy part of the proposed MAC-layer because aimed applications, like industrial sensor network, generally use embedded devices.